\documentstyle[12pt]{article}
\begin{document}

\begin{titlepage}
\title{An Effective Strong Gravity induced by QCD}
\author{Vincent Brindejonc\\
CEA,DAPNIA \\
Service de Physique Nucl\'eaire\\
\and Gilles Cohen-Tannoudji\\
CEA,DAPNIA \\
Service de Physique des Particules\\
CE Saclay 91191 Gif sur Yvette Cedex}
\date{}

\maketitle
\begin{abstract}
We show that, when quantized on a curved ``intra-hadronic background'',
QCD induces an effective pseudo gravitational interaction with gravitational
and cosmological constants in the GeV range.
\end{abstract}
\end{titlepage}

\section{Introduction}

In the pre-QCD phenomenology of strong interactions, several features
suggested that a strong gravity
\footnote{i.e. gravitational interaction with a coupling constant
$G_f$ $10^{38}$ times larger than the Newton's constant and, a spin two
mediator with vacuum quantum numbers identified as the $f_2$ meson.}
is at work in the hadronic world. In this phenomenology one has linearly
rising Regge trajectories which suggest that hadrons be interpreted as black
holes of the strong gravity (see ref \cite{sinha} for a wide review).
In addition to this, the success of the hadronic string picture suggests the
presence of a strong gravity in its closed string sector.

More recently, the subject was revived in a series of papers by Ne'eman
and Sija\v{c}ki who try to identify the strong gravity (which they call
a pseudo-gravity) in the infrared sector of QCD \cite{neeman}, and in a
paper by Salam and Sivaram \cite{salam}, connecting strong gravity with
confinement.

In the present letter, we intend to show that it is possible to derive the
strong
gravity as an effective theory induced by QCD when quantized in a ``curved
hadronic space-time'' (which we call ``aggravated QCD'').
To do this we use the method of Adler and Zee (see section \ref{SECADLERZEE}).
This method leads to an induced gravitational constant $G_{ind}$ and an induced
cosmological constant $\Lambda_{ind}$ which are in principle calculable using
the
parameters of the theory of matter quantized on a flat space-time.
Unfortunately, for a description of the true gravity, it seems difficult with
this method to obtain, for a given theory of matter, a gravitational constant
compatible with $G_{Newton}$ and a cosmological constant, which would not
violently contradict astrophysical observations.
When applying the Adler and Zee method to QCD quantized on this ``curved
hadronic
space-time'' it leads to a strong gravitation like interaction induced by QCD.
The large induced cosmological constant, which is a difficulty for true
gravity,
may be an advantage for strong gravity as it acts over a finite range.

\section{The method of Adler and Zee } \label{SECADLERZEE}

The idea of this method (see \cite{adler} for a complete review) was to derive
Einstein's gravity as an effective theory induced by the quantum fluctuations
of the matter fields. The fundamental Lagrangian is thus written as
\begin{equation}
\label{EQADLER}
{\cal L}={\cal L}^{cov}_{matter}(\Phi,g_{\mu\nu})+{\cal L}_{grav}(g_{\mu\nu})
\end{equation}
where $\Phi$ represents the matter fields, $g_{\mu\nu}$ the metric field,
${\cal L}^{cov}_{matter}$ the ``covariantized'' matter Lagrangian and
${\cal L}_{grav}$ the gravity Lagrangian.

If we consider ${\cal L}^{cov}_{matter}$ renormalizable, without bare masses
and ${\cal L}_{grav}$ purely quadratic in the curvature
\footnote{See equation (~\ref{EQLAGPSEUGRAV}) in Section ~\ref{SECAGRAVQCD}
below.},
then ${\cal L}$ is scale invariant. This invariance is dynamically broken by
renormalization. This fact renders calculable the parameters of an effective
low energy theory  generated by summation over matter fields.

The effective theory is defined by:
\begin{equation}
\label{EQEFFGRAV}
\exp(i{\cal S}_{eff}[g_{\mu\nu}]) =
\int{\cal D}\Phi\exp(i\int d^4x\sqrt{-det(g_{\mu\nu})}\,{\cal L}
(\Phi,g_{\mu\nu}))
\end{equation}
with
$${\cal S}_{eff}[g_{\mu\nu}] = \int{d^4x \sqrt{-det(g_{\mu\nu})}
{\cal L}_{eff}(g_{\mu\nu})}.$$
For slowly variable metrics, ${\cal L}_{eff}$ can be expanded in powers of
$\partial_\lambda g_{\mu\nu}$ so that,
\begin{equation}
\label{EQEXPDG}
{\cal L}_{eff}(g_{\mu\nu}) =
{\cal L}^{(0)}_{eff}(g_{\mu\nu}) +
{\cal L}^{(2)}_{eff}
(\partial_\lambda\partial_\tau g_{\mu\nu},(\partial_\lambda g_{\mu\nu})^2) +
O[(\partial_\lambda g_{\mu\nu})^4]\, .
\end{equation}
It can be identified with an Einstein-Hilbert Lagrangian, up to the
second order, if we write
\begin{equation}
\label{EQGLAMBIND}
{\cal L}^{(0)}_{eff}(g_{\mu\nu}) =
{1\over {8\pi}}{\Lambda_{ind}\over G_{ind}}\,\,\,\,\,
{\cal L}^{(2)}_{eff}(\partial_\lambda \partial_\tau g_{\mu\nu},
(\partial_\lambda g_{\mu\nu})^2) =
{1\over{16\pi G_{ind}}}R\, .
\end{equation} where $R$ is the scalar curvature built from $g_{\mu\nu}\,.$
Application of $\delta/\delta g^{\rho\sigma}$ on equation (\ref {EQEFFGRAV})
gives
\footnote{We follow the convention of \cite{gravitation} save for the metric
which has here signature $(+,-,-,-)$}
\begin{equation}
\label{EQEINSTEIN}
{1\over{16\pi G_{ind}}}(R_{\rho\sigma}-{1\over 2} R\, g_{\rho\sigma}\, +
\,\Lambda_{ind}g_{\rho\sigma}) = {1\over 2}<T_{\rho\sigma}>_{g,0}
\end{equation}
where we have defined :
\begin{equation}
{1\over 2}<T_{\rho\sigma}>_{g,0} = { 1\over{ \sqrt{-det(g_{\mu\nu})} } }
\frac{ \int {\cal D} \Phi {\delta\over {\delta g^{\rho\sigma}}}
S_{matter}^{cov}
\exp(iS) } {\int{\cal D}\Phi \exp(iS)}
\end{equation}
where the notation $<\,>_{g,0}$ indicates that the mean value is taken in the
background $(g_{\mu\nu})$ and in the vacuum $(0)$.
We will see that equation (\ref{EQEINSTEIN}) can be used under certain
conditions
to determine ${\Lambda_{ind}\over G_{ind}}$ and ${1\over{16\pi G_{ind}}}$.
Through dynamical scale breaking in flat space-time, eq.(\ref{EQEINSTEIN})
leads
to \cite{adler}:

\begin{equation}
\label{EQLAMBDASURG}
{\Lambda_{ind}\over {2\pi G_{ind}}} =
<T_\mu^\mu>_{\eta,0} = {\beta(g)\over 2g} <F^a_{\mu\nu}F_a^{\mu\nu}>_{\eta,0}
\end{equation}

\begin{equation}
\label{EQGIND}
{1\over{16\pi G_{ind}}} =
{i\over 96}{\int d^4x x^2(<\tau(T(x)T(0))>_{\eta,0}-<T(0)>_{\eta,0}^2)}
\end{equation}
where $\tau$ denotes the time ordered product.

We can pose the question as to wether the flat space-time approximation
is valid in this case.
In fact, splitting $g_{\mu\nu}$ into a background metric and a quantum
fluctuation, $g_{\mu\nu}=\overline{g_{\mu\nu}}+ \epsilon h_{\mu\nu}$,
this approximation means $\overline{g_{\mu\nu}} = \eta_{\mu\nu}$
and $\epsilon$ small.
It is justified at high energy scale, if on the one hand we do not consider
the presence of background matter fields and if the classical equations for
$\overline{g_{\mu\nu}}$ are solved in {\bf R}$^4$ without boundaries and on
the other hand, if the theory is asymptotically free
\footnote{As is the case both for an SU(N) gauge theory and for scale invariant
 gravitation \cite{fradkin}}.
As ${\Lambda_{ind}\over G_{ind}}$ and ${1\over{16\pi G_{ind}}}$ are
renormalisation group invariants, we can determine them for any energy scale
$\mu$ and in particular in the limit of large energy scales.

According to the phenomenon of dimensional transmutation (see the so-called
Gross-Neveu theorem \cite{gross}), the dimensionless coupling constant
is exchanged in favor of the mass gap of the theory, ${\cal M}(g(\mu),\mu)$.
${\cal M}(g(\mu),\mu)$ is solution of the renormalization group equation,
\begin{equation}
\label{EGR}
[\mu{\partial\over\partial\mu} +
\beta(g){\partial\over\partial g}]{\cal M}(g,\mu) = 0
\end{equation}
With equation (\ref{EGR}) one may express any physical quantity ${\cal P}$ of
canonical dimension $d_{\cal P}$, as
$${\cal P}=c{\cal M}^{d_{\cal P}}$$
where c is a dimensionless constant.
With this and from (\ref{EQLAMBDASURG}) one has immediately, for a $SU(N_c)$
gauge theory,
\begin{equation}
\label{LSURGFIN}
{\Lambda_{ind}\over {2\pi G_{ind}}} = -{2\pi\over 24}(11 N_c -2N_f)c{\cal M}^4
\end{equation}
where c is a numerical constant of the order of unity. The result for
(\ref{EQGIND}) is more difficult to obtain, in \cite{khury} upper and lower
bounds have been derived:
\begin{equation}
\label{UNSURGFIN}
{5\over{48 ln(10)}}(N_c^2-1){\cal M}^2\leq {1\over {16\pi G_{ind}}} \leq
{25\over{12 \pi^2}}(N_c^2-1){\cal M}^2
\end{equation}
It is clear when we see (\ref{LSURGFIN}) and (\ref{UNSURGFIN}) that this method
cannot be applied to true gravitation at least in this form because it gives
$\Lambda_{ind}$ of order of ${1\over G_{ind}}$.
It will however, be very useful when applied to a ``pseudo gravity'' associated
with the strong interaction.

\section{QCD in a curved intra-hadronic space-time}
\label{SECAGRAVQCD}

\subsection{The intra-hadronic background field}

We first remark that if one starts from standard QCD as the matter input,
the method of Adler and Zee leads to induced gravitational and cosmological
constants with values in the GeV range since ${\cal M}_{QCD}(g,\mu)$ is
itself in this range, and since QCD is asymptotically free.
Of course such an induced gravity would have nothing to do with the true
universal gravity, its metric field is certainly not the universal metric of
space-time.

We argue that due to the finite range of strong interactions, it is impossible
to probe with strongly interacting particles the structure of space-time inside
a hadron, in such a way that there is no compelling reason to assume this
space-time to be flat.
On the other hand, because of confinement, the only space-time seen by the
quarks
and the gluons is this intra-hadronic space-time.
We thus modify QCD by quantizing it on a ``curved intra-hadronic background''.

We treat $G_{\mu\nu}(x)$, the intra-hadronic metric field as:
(i) an {\em independant} field, namely a field which does not depend on the
matter fields,
(ii) a {\em specific} field, which means that the photon and the weak bosons
are coupled to quarks through the flat metric $\eta_{\mu\nu}$ and not through
$G_{\mu\nu}(x)$.

\subsection{The lagrangian of ``aggravated QCD''}

The Lagrangian of our theory, which we have named ``aggravated QCD'' can then
be
written as
\begin{equation}
\label{EQAGRQCD}
{\cal L}_{had} = {\cal L}_{PG}(G_{\mu\nu}) +
{\cal L}^{cov}_{QCD} (A^a_{\mu},\psi^{i,\alpha},G_{\mu\nu})
\end{equation}
where, the subscript ``$had$'' means that no non-hadronic field is involved,
 ${\cal L}_{PG}$ is the pseudo gravitational Lagrangian build from
 $G_{\mu\nu}(x)$ and ${\cal L}^{cov}_{QCD}$ is the standard QCD Lagrangian
 ``covariantized'' by means of $G_{\mu\nu}(x)$. In order to ``covariantize''
a Lagrangian involving spinors it is usefull to decompose the metric
$G_{\mu\nu}(x)$ in terms of the tetrads $e^m_{\mu}$
\begin{equation}
\label{EQTETRAD}
G_{\mu\nu} = e^m_\mu\,\eta_{mn}\,e^n_\nu
\end{equation}
where the latin indices ($m,n,\dots$) correspond to coordinates in the tangent
space ($M_4$) and $\eta_{mn}$ is the Minkowski metric in $M_4$, $\eta_{mn}$ =
diag(+1,-1,-1,-1).
We thus write
\begin{equation}
\label{EQLQCDCOV}
{\cal L}^{cov}_{QCD}(A^a_{\mu},\psi^{i,\alpha},G_{\mu\nu}) = -
{1\over4}F^a_{\mu\nu}F^a_{\rho\sigma}\, {1\over2}(G^{\mu\rho}\,
G^{\nu\sigma}-G^{\mu\sigma}G^{\nu\rho}) + {\bar \psi}_{i,\alpha}
\gamma^\mu{\stackrel{\leftrightarrow} {\nabla}}_\mu\psi^{i,\alpha}
\end{equation}
where
\begin{itemize}
\item the gluon fields are denoted by $A^a_\mu$, $a=1,\dots,N_c^2-1$, ($N_c$
is the number of colors),
\item the massless quark fields are denoted by
$\psi_{i,\alpha}$, $i=1,\dots,N_c$, $\alpha=1,\dots,N_f$ ($N_f$ is the
number of quark flavors),
\item $F^a_{\mu\nu}=\partial_\mu\,A^a_\nu-\partial_\nu\,A^a_\mu\,+
igf^a_{\;bc}A^b_\mu A^c_\nu$, ($f^a_{\;bc}$ are the structure constants
of the $SU(N_c)$ group), $g$ is the bare QCD coupling constant,
\item
\begin{equation}
\label{EQDECOV}
\gamma^\mu{\stackrel{\leftrightarrow}{\nabla}}_\mu =
e^\mu_m[\gamma^m(\stackrel{\leftrightarrow}{\partial}_\mu -
igA^a_\mu\tau_a)\,+{i\over2} \Gamma_{n\mu p}\{\gamma^m;\Sigma^{np}\}],
\end{equation}
where
\begin{itemize}
\item $\stackrel{\leftrightarrow}{\partial}_\mu =
\stackrel{\leftarrow}{\partial}_\mu\,- \stackrel{\rightarrow}{\partial}_\mu$,
\item $\tau_a$ are the generators of $SU(N_c)$,
\item $\Gamma_{n\mu p}$ are the Cristoffel symbols associated with
$G_{\mu\nu}$,
\item $\gamma^m$ are the Dirac matrices.
\item $\Sigma^{np}={i\over2}[\gamma^n,\gamma^p]$,
\end{itemize}
\end{itemize}

In order to preserve scale invariance, which holds if quarks are massless,
the pseudo-gravitational Lagrangian is purely quadratic in the curvatures:

\begin{equation}
\label{EQLAGPSEUGRAV}
{\cal L}_{PG}(G_{\mu\nu}) =
\alpha\,R_{\mu\nu\rho\sigma}R^{\mu\nu\rho\sigma}
 + \beta\,R_{\mu\nu}R^{\mu\nu} +
\gamma\,R^2
\end{equation}
where $R^\mu_{\;\nu\rho\sigma}$, $R_{\mu\nu}$ and $R$ are respectively the
Riemann tensor, the Ricci tensor and the scalar curvature associated with
the metric field $G_{\mu\nu}(x)$, and where $\alpha$, $\beta$ and $\gamma$
are dimensionless coupling constants.

Such a pseudo-gravitational Lagrangian has many advantages for our purpose.
(i) It ensures renormalizability \cite{stelle} and
(ii) asymptotic freedom \cite{fradkin} for the pseudo-gravitational
interaction.
(iii) It leads to a ${1\over p^4}$ high momentum behavior for the
pseudo-graviton
propagator in such a way that we can expect aggravation not to modify the
$\beta(g)$ function of QCD.
The price to pay these good features is the possible occurence of ghosts at
least
at the perturbative level.

\subsection{Symmetry properties of aggravated QCD}

By construction, aggravation adds a symmetry to standard QCD, namely ``general
hadronic covariance'', the invariance by a general local change of the
intra-hadronic coordinates at which the quark and gluon fields are defined,
whereas it preserves its symmetries. In fact since $G_{\mu\nu}$ does not depend
on the QCD fields, the aggravation of QCD preserves the $SU(N_c)$ gauge
invariance.
This feature has to be contrasted with the breaking of gauge invariance implied
by the metric used in ref. \cite{neeman} and \cite{salam}. The fact that the
photon and the weak bosons couple to quarks through $\eta_{\mu\nu}$ and not
through $G_{\mu\nu}$ means that electromagnetic and weak interactions break
general hadronic covariance, a specific symmetry of strong interactions in the
same way as they break isospin invariance, another specific symmetry of strong
interactions.

It turns out that chiral symmetry, a property of QCD with massless quarks, is
also preserved by our aggravation procedure.
To see this it is enough to note that the only term which may pose a problem,
namely the fermionic term in (\ref{EQLQCDCOV}), involves, as shown in
(\ref{EQDECOV}), only odd numbers of Dirac matrices and is thus chiral
invariant:
\begin{equation}
\label{EQCHIRSYM}
{\bar\psi}\gamma^\mu\stackrel{\leftrightarrow}{\nabla}_\mu\psi =
{\bar\psi_R}\gamma^\mu\stackrel{\leftrightarrow}{\nabla}_\mu
\psi_R+{\bar\psi_L}\gamma^\mu\stackrel{\leftrightarrow}{\nabla}_\mu \psi_L
\end{equation}
where the eigenvectors of chirality are defined by
\begin{equation}
\label{EQCHIREIG}
\psi_{\stackrel{R}{L}} = {1\pm\gamma_5\over2}\psi
\end{equation}
As in ordinary QCD with massless quarks, chiral symmetry is thus broken only
dynamically in aggravated QCD.
Nevertheless, since the conservation of chiral symmetry currents must be
written
in terms of the covariant derivative i.e. $\partial_{\mu}(\sqrt{-G}J^\mu)=0$,
we
can have departures from standard chiral theory.
However, on the one hand, the ultra-violet decoupling of pseudo-gravity
guarantees that we recover the standard formula at high energy, and on the
other hand, the effective metric which we propose below -see
eq.(\ref{EQSOLPEINSTEIN})-  does not contribute to $\sqrt{-G}\,\,\,$
\footnote{except through the standard spheric term}.
In general, as in (\ref{EQSOLPEINSTEIN}), we expect pseudo-gravity not to
contribute at energies lower than ${\cal M} \equiv \Lambda_{QCD}$, and thus
to preserve one of the main assets of standard QCD, namely chiral perturbation
theory.

\section{The induced Strong Gravity}

The Lagrangian of aggravated QCD satisfies all the criteria (renormalizability,
scale invariance, asymptotic freedom, absence of bare Einstein-Hilbert terms)
to lead through the method of Adler and Zee to induced gravitational and
cosmological constants which are finite and in principle calculable in terms of
the matter theory quantized on a flat space-time. One can thus say that
aggravated QCD induces an effective strong gravity.
This strong gravity is characterized by its metric field $G_{\mu\nu}$, by its
strong gravitational constant $G_{f}=G_{ind}$ and its strong cosmological
constant $\Lambda_{f}=\Lambda_{ind}$.
With $N_c=3$ and ${\cal M}=\Lambda_{QCD}\sim 200MeV$, we find that $G_f$ and
$-\Lambda_f$
\footnote{with our conventions, the induced cosmological constant is negative}
in the GeV range.

A first application of our scheme consists on trying to find the metric of the
effective pseudo-gravitation induced by aggravated QCD. To this end one has to
consider eq.(\ref{EQEINSTEIN}), in which $g_{\mu\nu}$ is replaced by
$G_{\mu\nu}$.
Despite its appearance, this equation should be understood as a definition of
an
effective long distance energy-momentum tensor rather than as the Einstein
equation itself. We thus can write
\begin{equation}
\label{EQPEINSTEIN}
<T_{\rho\sigma}>_{G,0}\equiv T_{\rho\sigma}^{eff} =
{1\over{8\pi G_{ind}}} (R_{\rho\sigma} - {1\over 2} R\,G_{\rho\sigma}\,+
\,\Lambda_{ind}\,G_{\rho\sigma})
\end{equation}
If one wants to obtain the intra hadronic metric induced by the quantum
fluctuations of the standard matter theory, one has to equal to zero the
full (QCD plus pseudo-gravity) effective energy-momentum tensor, in such
a way that eq.(\ref{EQPEINSTEIN}) can now be solved as an Einstein
equation in the vacuum.
This equation leads to a closed space-time since $\Lambda_{ind}$ is negative.
In this situation, the ``non abelian Gauss theorem'' implies that the only
physical states are color-singlet. For instance, if the vacuum fluctuations
which have been integrated on, have been created by a quark source situated
at a given point of space-time, then there necessarily exists an antiquark
(or a diquark) in the light cone of the quark on which the color field lines
must reconverge.

The solutions of the vacuum equation involve a de Sitter intra-hadronic
space-time. The metric can take several forms, including, in a comoving
frame, the one of an ever expanding space-time. However it can be useful,
for our purpose, to interpret the solution as it can be seen by an extra
hadronic probe which couples to quarks through $\eta_{\mu\nu}$. We obtain
this way the static solution of the vacuum equation
\begin {equation}
\label{EQSOLPEINSTEIN}
ds^2 =
(1+{1\over3}\Lambda_{ind}r^2)dt^2 -
(1+{1\over3}\Lambda_{ind}r^2)^{-1}dr^2 -
r^2d\theta^2 -
r^2\sin^2\theta d\phi^2
\end{equation}
This metric is flat at short distance (in accordance with asymptotic freedom)
and presents a singularity at $r_s=\sqrt{-3\over\Lambda_{ind}}$ which can be
interpreted as the radius of a ``gravitational bag''. In this metric, the
equation of an outgoing null radial geodesic is
\begin{equation}
\label{EQOUTNURAGEO}
{t-t_0\over r_s} =
{1\over 2}\ln \left |{1+r/r_s\over1-r/r_s}\right |
\end{equation}
showing that the chromo-electric field produced by a quark at the origin never
reaches $r=r_s$ and is thus confined at a distance smaller than $r_s$. We note
that this solution is not a black hole; it is rather the exterior of the bag
which resembles the interior of a Schwartchild black hole. The anti-screening
effect of vacuum polarization in QCD (see \cite{lee}) is simulated in terms of
a strong gravitational effect.

This encouraging result has to be considered as a first stage for a better
understanding of the pseudo-gravity induced by QCD.
Much more work is needed (and in progress), in particular in the study of the
``fundamental'' lagrangian and of the modifications of QCD implied by our
aggravation procedure, and in the construction of effective theories with
background quark and gluon fields. One of the first applications of aggravated
QCD could be to relate its metric field $G_{\mu\nu}(x)$ to the so called ``hard
Pomeron'' which is advocated in the interpretation of the recent Hera data
\cite{hera}.

\vskip 0.8cm
\noindent{\bf Aknowledgments}
\noindent It is a pleasure to thank several stimulating discussions with
Jean-Claude Le Guillou, Marc Knecht, Jean-Fran\c cois Picard, Marion
Mac Cormick and P. Teyssandier.

{}

\end{document}